\documentclass[11pt]{article}
\usepackage{bbm}
\usepackage{amsfonts}
\usepackage{mathrsfs}
\usepackage{pifont}
\usepackage{amsmath}
\usepackage{amssymb}
\usepackage{caption2}
\usepackage{graphicx}
\begin{document}
\title{Lower Bound of Concurrence for Qubit Systems}
\author{ Xue-Na Zhu$^{1}$
 \ \ \ Shao-Ming Fei$^{2,3}$ \\
{\footnotesize  {$^1$Department of Mathematics,
 South China University of Technology, Guangzhou
510640, P.R.China}} \\
{\footnotesize{
  $^2$School of Mathematical Sciences, Capital Normal University,
Beijing 100048, China}}\\
{\footnotesize{$^3$Max-Planck-Institute for Mathematics in the Sciences, 04103
Leipzig, Germany}}}

\date{}
\maketitle

{\bf Abstract:}
We study the concurrence of four-qubit quantum states and provide
analytical lower bounds of concurrence in terms of the monogamy inequality of concurrence for qubit systems.
It is shown that these lower bounds are able to improve the existing bounds and detect
entanglement better. The approach is generalized to arbitrary qubit systems.

\section*{1. Introduction}
Quantum entanglement plays important roles not only in quantum information science
\cite{s2,s3,s4,s5}, but also in many fascinating features in quantum theory, which have puzzled generations of physicists \cite{Amico}.
The fundamental problems in the theory of quantum entanglement is the entanglement detection and quantification.
Concurrence \cite{s6,s7,s8} is an important measure of quantum entanglement \cite{t1,t2,t3,t4,t5,t6}.
Different from entanglement of formation which works only for bipartite systems \cite{e11,e12}, concurrence
can be generalized to arbitrary multipartite systems.
However, due to the extremizations involved in the computation,
analytical formulas of concurrence are available only for two-qubit states \cite{PRL2245} and some high dimensional
bipartite states with certain symmetries, like isotropic
states and Werner states \cite{PRA062307} and some special symmetric states \cite{Terhal-Voll2000,Terhal-Voll20002,Terhal-Voll20003}.
Calculation of concurrence for general quantum states is a formidable task.
In particular, quite less has been known about the concurrence of multipartite mixed states.

In \cite{Xiu-hong Gao.2006} analytical lower bounds of concurrence for three-qubit states
have been presented based on PPT (positive partial transposition) and realignment operations.
Lower bounds of concurrence for $M$-qubit states with pure state decompositions
given by the generalized Greenberger-Horne-Zeilinger states or the generalized $\textrm{W}$-state,
and for the multipartite $SC$ (Schmidt correlated) states are provided in \cite{Xiu-hong Gao.2008}.
Analytical lower bounds of concurrence for general multipartite systems have been discussed in terms of all possible
bipartite decompositions in \cite{Front}.

In this paper, by using a new approach we provide an analytical lower
bound of concurrence for general four-qubit mixed quantum states based on monogamy inequalities.
The results are generalized to multipartite case.
Our lower bounds can improve the previous ones in Refs. \cite{Front,PRA062320}.

\section*{2. The lower bounds of concurrence for four-qubit systems}
Let $H_{1}$, $H_{2}$, ..., $H_{N-1}$ and  $H_{N}$ be $N$ $2$-dimensional
vector spaces associated with $N$ quantum systems.
The concurrence of a state ${\left\vert \psi \right\rangle }\in H_1\otimes H_2\otimes...\otimes H_N$ is
defined by, up to an $N$ dependent factor $2^{1-\frac{N}{2}}$ \cite{t1,prl93},
\begin{equation}\label{NCpure}
C_{12...N}({\left\vert \psi \right\rangle })=\sqrt{2^{N}-2-\sum_{\alpha}Tr\rho^{2}_{\alpha}},
\end{equation}
where the index $\alpha$ labels all $2^N-2$
non-trivial subsystems of the $N$-qubit
system and $\rho_{\alpha}$ are the the corresponding  reduced density matrices \cite{prl97}.

The concurrence for a mixed state $\rho$ is defined by the convex roof,
\begin{equation}\label{e1}
C(\rho )\equiv \min_{\{p_{i},|\psi _{i}\rangle
\}}\sum_{i}p_{i}C({\left\vert \psi _{i}\right\rangle }),
\end{equation}
for all possible pure state decompositions $\rho
=\sum_{i}p_{i}|\psi_{i}\rangle\langle \psi _{i}|$, where
${\left\vert \psi_i \right\rangle } \in {\mathcal H}_{1} \otimes
{\mathcal H}_{2}\otimes...\otimes {\mathcal H}_{N}$, $0\leq p_{i}\leq1$ and
$\sum_{i}p_{i}=1$.

For two-qubit case, the concurrence of a two-qubit state $\rho\in H_1\otimes H_2$ is given by
\begin{equation}\label{c2}
C(\rho)=\max\{\lambda_1-\lambda_2-\lambda_3-\lambda_4,0\},
\end{equation}
with $\lambda_1\geq\lambda_2\geq\lambda_3\geq\lambda_4$ the square roots of the four nonzero eigenvalues
of the non-Hermitian matrix $\rho\tilde{\rho}$, $\tilde{\rho}=(\sigma_y\otimes\sigma_y)\rho^{*}(\sigma_y\otimes\sigma_y)$,
where ${*}$ denotes complex conjugation in the standard basis and $\sigma_y$ is the Pauli matrix.

For convenience, we define bracket $\{a|b\}$. One may either take the first element $a$ or the second element
$b$ from $\{a|b\}$. However, for any given pair $a$ and $b$, once the first (the second) has been taken, then
in a formula one always takes the first (the second) element in all the following brackets containing the same two elements $a$ and $b$.
Namely, if one takes $\{a|b\}=a$, then $\{b|a\}=b$; or if one takes $\{a|b\}=b$ then $\{b|a\}=a$. We set
\begin{equation}
T_1=\displaystyle 1+\{-\frac{2-x}{2}|\frac{2-x}{2}\}+\{-\frac{2-y}{2}|\frac{2-y}{2}\}+ \{-\frac{2-z}{2}|\frac{2-z}{2}\},
\end{equation}
\begin{equation}
T_2=1+\{\frac{2-x}{2}|-\frac{2-x}{2}\}+\{-\frac{y}{2}|\frac{y}{2}\}+\{-\frac{z}{2}|\frac{z}{2}\},
\end{equation}
 \begin{equation}
 T_3=1+\{-\frac{x}{2}|\frac{x}{2}\}+\{\frac{2-y}{2}|-\frac{2-y}{2}\}+\{\frac{z}{2}|-\frac{z}{2}\},
\end{equation}
and
\begin{equation}
T_4=1+\{\frac{x}{2}|-\frac{x}{2}\}+\{\frac{y}{2}|-\frac{y}{2}\}+\{\frac{2-z}{2}|-\frac{2-z}{2}\},
\end{equation}
where $x,y,z\in[0,2]$. $T_i$, $i=1, 2, 3, 4$ are all greater than or equal to zero for all choices
of $x, y$ and $z$ and for each of the two choices allowed by our bracket notation.

For $N$-qubit quantum states the concurrence satisfies the monogamy inequality \cite{PRL220503}:
\begin{equation}\label{monogamy}
C^2_{A_1|A_2A_3...A_N}(\rho)\geq \sum_{i=2}^{N}C^2_{A_1A_i}(\rho),
\end{equation}
where $C_{A_1|A_2A_3...A_N}(\rho)$ is the concurrence of state $\rho$
under the bipartite bipartition $A_1$ and $A_2A_3...A_N$, and $C_{A_1A_i}(\rho)$ denotes
the concurrence of the reduced state $\rho_{A_1A_i}=Tr_{A_2...A_{i-1}A_{i+1}...A_{N}}(\rho)$, $i=2,...,N$.
We denote $C_{i|jkl}$ (resp. $C_{ij|kl}$) the bipartite concurrence under the
bipartition $i$ and $jkl$ (resp. $ij$ and $kl$), where $i\not=j\not=k\not= l\in\{1,2,3,4\}$.

{\bf Theorem 1:}\label{T1}
For any four-qubit mixed quantum
state $\rho$, the concurrence $C(\rho)$ satisfies
\begin{equation}\label{xia}
C^2(\rho) \geq \sum_{i=1}^{3}\sum_{j>i}^{4}(T_i+T_j)C^2_{ij}(\rho).
\end{equation}

[Proof:]
The concurrence (\ref{NCpure}) of a four-qubit pure state $|\psi\rangle$ can be equivalently written as
\begin{equation}\label{4Cpure}
C^2_{1234}({\left\vert \psi \right\rangle })=\displaystyle 2\sum_{i=1}^{4}(1-Tr(\rho^2_i))
+2[(1-Tr^2(\rho_{12}))
+(1-Tr^2(\rho_{13}))+(1-Tr^2(\rho_{14}))].
\end{equation}
From (\ref{4Cpure}) one has
\begin{equation}\label{c1234}
C^2_{1234}({\left\vert \psi \right\rangle })
=C^2_{1|234}+C^2_{2|134}+C^2_{3|124}+C^2_{4|123}
+C^2_{12|34}+C^2_{13|24}+C^2_{14|23}.
\end{equation}

The bounds of the terms $C^2_{12|34}$ $C^2_{13|24}$ and $C^2_{14|23}$ in
(\ref{c1234}) can be further derived.
Since $Tr(\rho^2_{12})=Tr(\rho^2_{34})$ for a four-qubit pure state $|\psi\rangle$, we have
$$
C^2_{12|34}({\left\vert \psi \right\rangle })=x(1-Tr(\rho^2_{12}))+(2-x)(1-Tr(\rho^2_{34})),
$$
where $x\in[0,2]$. Therefore
\begin{equation}\label{c12}
C^2_{12|34}({\left\vert \psi \right\rangle })\geq x[(1-Tr(\rho^2_{12p}))-(1-Tr(\rho^2_p))]
+(2-x)[(1-Tr(\rho^2_{34q}))-(1-Tr(\rho^2_q))]
\end{equation}
for $p\in\{3,4\}$ and $q\in\{1,2\}$, where
the relation $1+Tr(\rho^2_{AB})\geq Tr(\rho^2_A)+Tr(\rho^2_B)$ in
\cite{PRA042308} for bipartite states $\rho_{AB}$ has been used.
Four different combinations of choosing $p$ and $q$ in (\ref{c12}) give rise to that
$C^2_{12|34}({\left\vert \psi \right\rangle })$ is greater or equal to the following
four formulae:
$$
\frac{x}{2}\left(C^2_{4|123}-C^2_{3|124}\right)
+\frac{2-x}{2}\left(C^2_{2|124}-C^2_{1|123}\right),
$$
$$
\frac{x}{2}\left(C^2_{4|123}-C^2_{3|124}\right)
+\frac{2-x}{2}\left(C^2_{1|123}-C^2_{2|124}\right),
$$
$$
\frac{x}{2}\left(C^2_{3|124}-C^2_{4|123}\right)
+\frac{2-x}{2}\left(C^2_{2|124}-C^2_{1|123}\right)
$$
and
$$
\frac{x}{2}\left(C^2_{3|124}-C^2_{4|123}\right)
+\frac{2-x}{2}\left(C^2_{1|123}-C^2_{2|124}\right).
$$
For simplicity, we write
\begin{equation}\label{x}
C^2_{12|34}({\left\vert \psi \right\rangle })\geq  \displaystyle\frac{x}{2}\left\{C^2_{4|123}-C^2_{3|124},C^2_{3|124}-C^2_{4|123}\right\}
+\displaystyle\frac{2-x}{2}\left\{C^2_{2|124}-C^2_{1|123},C^2_{1|123}-C^2_{2|124}\right\},
\end{equation}
where $\{a,b\}$, different from the definitions of $\{|\}$, could be either $a$ or $b$.

Similarly we have
\begin{equation}\label{y}
C^2_{13|24}({\left\vert \psi \right\rangle })\geq \displaystyle\frac{y}{2}\left\{C^2_{4|123}-C^2_{2|134},C^2_{2|134}-C^2_{4|123}\right\}
+\displaystyle\frac{2-y}{2}\left\{C^2_{3|124}-C^2_{1|234},C^2_{1|234}-C^2_{3|124}\right\}
\end{equation}
and
\begin{equation}
\label{z}
C^2_{14|23}({\left\vert \psi \right\rangle })\geq \displaystyle\frac{z}{2}\left\{C^2_{3|124}-C^2_{2|134},C^2_{2|134}-C^2_{3|124}\right\}
+\displaystyle\frac{2-z}{2}\left\{C^2_{4|123}-C^2_{1|234},C^2_{1|234}-C^2_{4|123}\right\},
\end{equation}
where $y,z\in[0,2]$.
Denote $T_{ij}=T_i+T_j$, from (\ref{c1234}), (\ref{x}), (\ref{y}) and (\ref{z}) we obtain
\begin{equation}
\begin{aligned}\label{1|3}
C^2_{1234}({\left\vert \psi \right\rangle })
&\geq T_1C^2_{1|234}+T_2C^2_{2|134}+T_3C^2_{3|124}+T_4C^2_{4|123}\\
&\geq (T_1+T_2)C^2_{12}+(T_1+T_3)C^2_{13}+(T_1+T_4)C^2_{14}\\
&~~~+(T_2+T_3)C^2_{23}+(T_2+T_4)C^2_{24}+(T_3+T_4)C^2_{34}\\
&=T_{12}C^2_{12}+T_{13}C^2_{13}+T_{14}C^2_{14}+T_{23}C^2_{23}+T_{24}C^2_{24}+T_{34}C^2_{34},
\end{aligned}
\end{equation}
where  the monogamy inequality (\ref{monogamy}) has been
used in the second inequality.

Let $\rho=\sum_ip_i|\psi\rangle_i\langle\psi|$ be the optimal pure state decomposition of (\ref{e1})
for a four-qubit mixed state $\rho$.
For any pure state $\rho^i=|\psi\rangle_i\langle\psi|$ in the decomposition,
we take the same parameters ${x,y,z}$ and the same way in choosing $\{a|b\}$ from $T_i\geq0$, $i=1,2,3,4$.
Denote $\beta$ the index set $\{12,13,14,23,24,34\}$. We have
\begin{equation}
\begin{aligned}C^2(\rho)
&=\displaystyle\left\{\sum_ip_iC(|\psi\rangle_i\langle\psi|)\right\}^2\\
&\geq\displaystyle\left\{\sum_ip_i\sqrt{(\sum_{\beta}T_{\beta}C^2_{\beta}(\rho^i))}\right\}^2\\
&\geq\displaystyle\sum_{\beta}(\sum_ip_i\sqrt{T_{\beta}}C_{\beta}(\rho^i))^2\\
&=\displaystyle\sum_{\beta}T_{\beta}(\sum_{i}p_iC_{\beta}(\rho^i))^2\\
&\geq\displaystyle\sum_{\beta}T_{\beta}C^2_{\beta}(\rho),
\end{aligned}
\end{equation}
where the relation $(\sum_j (\sum_i x_{ij})^2 )^{\frac{1}{2}} \leq
\sum_i (\sum_j x_{ij}^2)^{\frac{1}{2}}$ has been used in the second
inequality.
\hfill $\Box$

As there are free parameters $x$, $y$ and $z$, and many ways to
choose the elements in $T_i$, inequality (\ref{xia}) gives a set of lower bounds of the concurrence.
For example, we may fix $x=2$, $y=0$, $z=0$ and select appropriate combinations for $T_i$:
$T_1=1-\frac{2-x}{2}-\frac{2-y}{2}+\frac{2-z}{2}=1$,
$T_2=1+\frac{2-x}{2}-\frac{y}{2}-\frac{z}{2}=1$,
$T_3=1-\frac{x}{2}+\frac{2-y}{2}+\frac{z}{2}=1$, and
$T_4=1+\frac{x}{2}+\frac{y}{2}-\frac{2-z}{2}=1$.
Then we have  $C^2(\rho)\geq2[C^2_{12}+C^2_{13}+C^2_{14}+C^2_{23}+C^2_{24}+C^2_{34}]$.

If we denote $\Lambda$ the set of the lower bound of four-qubit states. In fact,
by taking suitable values of $x$, $y$ and $z$, and selecting appropriate combinations for $T_i (i=1,...,4)$, we have

\begin{equation}\left\{2[C^2_{i_1i_2}+C^2_{i_1i_3}+C^2_{i_1i_4}+C^2_{j_1j_2}+C^2_{j_1j_3}+C^2_{j_1j_4}],2[C^2_{12}+C^2_{13}+C^2_{14}+C^2_{23}+C^2_{24}+C^2_{34}]
\right\}\subseteq\Lambda
\end{equation}
where$\{i_1,i_2,i_3,i_4\}=\{j_1,j_2,j_3,j_4\}=\{1,2,3,4\}$.

To investigate the strength of the inequality (\ref{xia}), let us consider the following examples.

{\it Example 1.} We first consider a simple pure state,
$|\psi\rangle=|\psi^{+}\rangle\otimes|0\rangle\otimes|1\rangle$,
where $|\psi^+\rangle=({|00\rangle+|11\rangle})/{\sqrt{2}}$.
From the Eq.(\ref{4Cpure}), we have $C_{1234}(|\psi\rangle)=2$.
For this state, one has
$C_{12}(|\psi\rangle)=1$ and $C_{13}(|\psi\rangle)=C_{14}(|\psi\rangle)=0$.
we take $x=y=z=0$, and
$T_1=1+\frac{2-x}{2}+\frac{2-y}{2}+\frac{2-z}{2}=4$,
$T_2=1-\frac{2-x}{2}-\frac{y}{2}-\frac{z}{2}=0$,
$T_3=1-\frac{x}{2}-\frac{2-y}{2}+\frac{z}{2}=0$, and
$T_4=1+\frac{x}{2}+\frac{y}{2}-\frac{2-z}{2}=0$.
Then from the lower bound (\ref{xia}), we have
$C_{1234}^2(|\psi\rangle)\geq 4\left(C^2_{12}(|\psi\rangle)+C^2_{13}(|\psi\rangle)+C^2_{14}(|\psi\rangle)\right)=4$,
namely, the state $|\psi\rangle$ saturates the inequality (\ref{xia}).
Nevertheless, from the lower bound in \cite{Front} one has $C_{1234}(|\psi\rangle)\geq1$.
Hence our bound is better than the one given in \cite{Front}.

{\it Example 2.}
Let us consider the one-parameter four-qubit state
$$
\rho=\frac{1-a}{16}I_{16}+a|\psi\rangle\langle\psi|,
$$
where $|\psi\rangle=({|0011\rangle+|0101\rangle+|0110\rangle+|1010\rangle})/{2}$ and $I_{16}$ is the $16\times 16$ identity matrix.
From (\ref{c2}),
we get
$$
C_{12}(\rho)=\max\{\frac{1}{4}(\sqrt{1+a+a^2+2a\sqrt{1+a}}
-\sqrt{1+a+a^2-2a\sqrt{1+a}}-2\sqrt{1-a}),0\}.
$$
From the lower bound in \cite{PRA062320} $\rho$ is entangled
for $a>0.636364$. While if we take the same $x,y,z$ and $T_i$ as in example 1,
from (\ref{xia}) we have that $C(\rho)\geq2C_{12}(\rho)>0$ for $a>0.618034$. Hence
the bound (\ref{xia}) detects entanglement better.

For multipartite quantum systems, although there are some criteria to detect genuine multipartite entanglement,
there is no computable measure in quantifying the multipartite entanglement in general. The example below shows that
our lower bound of concurrence for multipartite quantum systems has a tight analytic
relations with two-qubit concurrences.

{\it Example 3.} We consider the quantum state
$\rho=\frac{1-t}{16}I_{16}+t|\psi\rangle\langle\psi|$,
where $|\psi\rangle=({|0000\rangle+|0011\rangle+|1100\rangle+|1111\rangle})/{2}$.
We have
$$
\rho_{12}=\displaystyle\rho_{34}=\frac{1+t}{4}\left(|00\rangle\langle00|+|11\rangle\langle11|\right)
+\displaystyle\frac{t}{2}\left(|00\rangle\langle11|+|11\rangle\langle00|\right)
+\frac{1-t}{4}\left(|01\rangle\langle01|+|10\rangle\langle10|\right),
$$
and
$$
\rho_{13}=\rho_{14}=\rho_{23}=\rho_{24}
=\displaystyle\frac{1}{4}(|00\rangle\langle00|+|01\rangle\langle01|+|10\rangle\langle10|+|11\rangle\langle11|).
$$
Therefore, by using the formula of concurrence for two-qubit states (\ref{c2}), we have
$$
C_{12}(\rho)=C_{34}(\rho)=\max\left\{0,\frac{\sqrt{1+6t+9t^2}-3(1-t)}{4}\right\}
$$
and $C_{13}=C_{14}=C_{23}=C_{24}=0$.
If we take $x=0, y=z=2$, and $T_1=1+\frac{2-x}{2}+\frac{2-y}{2}+\frac{2-z}{2}=2$,
$T_2=1-\frac{2-x}{2}+\frac{y}{2}+\frac{z}{2}=2$,
$T_3=1-\frac{x}{2}-\frac{2-y}{2}-\frac{z}{2}=0$, and
$T_4=1+\frac{x}{2}-\frac{y}{2}-\frac{2-z}{2}=0$,
then from Theorem 1, the lower bound of concurrence
is given by:
$$
C^2(\rho)\geq4C^2_{12}(\rho)+2C^2_{13}(\rho)+2C^2_{14}(\rho)+2C^2_{23}(\rho)+2C^2_{24}(\rho).
$$
From Fig. 1, we see that the lower bound can detect entanglement of $\rho$ when $t>{1}/{3}$.
\begin{figure}[htpb]
\renewcommand{\captionlabeldelim}{.}
\renewcommand{\figurename}{Fig.}
\centering
\includegraphics[width=7cm]{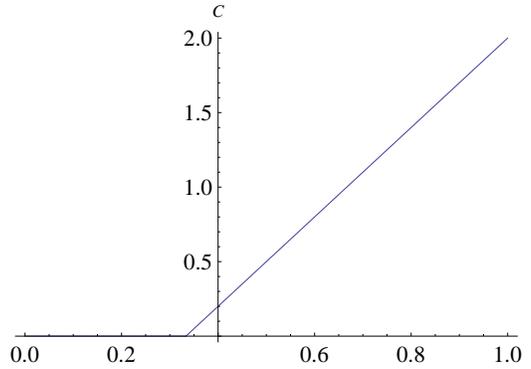}
\caption{{\small The lower bound concurrence of $\rho$ for $0\leq t\leq1$.}}
\label{2}
\end{figure}

\section*{3. the lower bounds of concurrence for arbitrary qubit systems}

Now we generalize our results to $N$-qubit systems.
For a given N-qubit state  ${\left\vert \psi \right\rangle }\in H_1\otimes H_2\otimes...\otimes H_N$,
the concurrence (\ref{NCpure}) has the form
$$
C^2_{12...N}(|\psi\rangle)=\sum_{\vec{j}}\frac{1}{2}\,C^2_{\vec{j}|R(\vec{j})},
$$
where $\vec{j}=\{j_1,j_2,...,j_r\}\subseteq\{1,2,...,N\}$ are all the possible
integer strings, $j_1<j_2<...<j_r$, such that $\vec{j}\cup R(\vec{j})=\{1,2,...,N\}$,
i.e. $R(\vec{j})=\{1,2,...,N\} \setminus \vec{j}$.

Similar to the four-qubit case, taking into account that $1+Tr{\rho^2_{AB}}\geq Tr(\rho^2_A)+Tr(\rho^2_B)$, we can prove the following corollary:

{\it Corollary 1:}\label{C1}
For any N-qubit pure state $|\psi\rangle$, the concurrence $C_{\vec{j}|R(\vec{j})}$ satisfies
\begin{equation}
C^2_{\vec{j}|R(\vec{j})}\geq\frac{1}{2}\left\{xC^2_{j_t|R(\{j_t\})}+(2-x)C^2_{j_s|R(\{j_s\})}
-x\sum_{j_p}C^2_{j_p|R(\{j_p\})}-(2-x)\sum_{j_q}C^2_{j_q|R(\{j_q\})}\right\},
\end{equation}
where: $j_t\in R(\vec{j})$, $j_s\in\vec{j}$, $j_p\in R(\vec{j})\setminus\{j_t\}$, $j_q\in\vec{j}\setminus\{j_s\}$ and $x\in[0,2]$.

From the  corollary, in terms of the monogamy relation (\ref{monogamy}),
for any N-qubit ($N\geq4$) mixed quantum
state $\rho$, there are some fixed numbers numbers $F_i\geq0$, $i=1,...,N$, which depend on parameters like $T_i$ in Theorem 1
and the concurrence $C(\rho)$ satisfies
\begin{equation}\label{fxia}
C^2(\rho) \geq \sum_{i=1}^{N-1}\sum_{j>i}^{N}(F_i+F_j)C^2_{ij}.
\end{equation}

\section*{4. Conclusion }
In summary, we have proposed a new approach in constructing hierarchy of lower
bounds of concurrence for four-qubit mixed states in terms of the
monogamy inequality of concurrence.
The lower bounds may be used to improve the previous lower
bounds of concurrence and can detect better quantum entanglement.
Besides, our approach can be generalized to $N-$qubit systems
to obtain the lower bound of the concurrence for $N$-qubit states.

\end{document}